\documentstyle[12pt]{article}
\def\cR{{\cal R}}
\def\caL{{\cal L}}
\def\a{\alpha}

\def\d{\delta}

\def\ga{\gamma}
\def\la{\lambda}

\def\m{\mu}
\def\n{\nu}
\def\r{\rho}
\def\s{\sigma}

\def\e{\eta}

\def\ee{\varepsilon}
\def\gm{\Gamma}

\def\La{\Lambda}
\def\parslash{{\partial{\hspace{-7pt}}/}}
\def\Aslash{{A{\hspace{-7pt}}/}}
\def\pslash{{p{\hspace{-5pt}}/}}
\def\kslash{{k{\hspace{-5.5pt}}/}}

\def\c{{\rm c}}
\def\f{{\rm f}}
\def\idk{\int\! {d^4\!k \over (2\pi)^4} \,\,}
\def\idx{\int\! d^4\!x \,}
\def\idy{\int\! d^4\!y \,}
\def\ds{\displaystyle}
\def\ss{\scriptstyle}

\def\eg{{\it e.g.}}
\def\to{\rightarrow}

\def\barc{\bar{c}}
\def\BB{\cal}
\begin{document}


\begin{titlepage}
\rightline{To appear in Phys. Lett. {\bf B355} (1995) 531}
\rightline{hep-ph/9507443}
\vskip 1 true cm
\begin{center}
  {\Large {\bf Unitarity violation in non-abelian\\
          \vskip 0.1 true cm
               Pauli-Villars regularization}}\\
  \vskip 1 true cm
  {\rm J.H. Le\'on}\\
  \vskip 0.2 true cm
  {\it Departamento de F\'\i sica Te\'orica, C-XI,
       Universidad Aut\'onoma de Madrid,} \\
  {\it Cantoblanco, 28049 Madrid, Spain}\\
  \vskip 1 true cm
  {\rm C.P. Martin}\\
  \vskip 0.2 true cm
  {\it Departamento de F\'\i sica Te\'orica I, Facultad de Ciencias
       F\'\i sicas,\\
       Universidad Complutense de Madrid, 28040 Madrid, Spain}\\
  \vskip 1 true cm
  {\rm F. Ruiz Ruiz}\\
  \vskip 0.2 true cm
  {\it NIKHEF-H, Postbus 41882, 1009 DB Amsterdam, The Netherlands}\\
\vskip 1.5 true cm
{\leftskip=.5 true cm \rightskip=.5 true cm
\noindent
We regularize QCD using the combination of higher covariant
derivatives and Pauli-Villars determinants proposed by Slavnov. It is
known that for pure Yang-Mills theory the Pauli-Villars determinants
generate unphysical logarithmic radiative corrections at one loop that
modify the beta function. Here we prove that when the gauge fields are
coupled to fermions so that one has QCD, these unphysical corrections
translate into a violation of unitarity. We provide an understanding
of this by seeing that Slavnov's choice for the Pauli-Villars
determinants introduces extra propagating degrees of freedom that are
responsible for the unitarity breaking. This shows that Slavnov's
regularization violates unitarity, hence that it should be rejected.
\par}
\end{center}
\end{titlepage}
\setcounter{page}{2}

\noindent

{\it 1 Introduction.} The advantages of using gauge invariant
regularization methods are well known to the quantum field theory
practitioner. The problem is that there are not so many gauge
invariant regularization methods available. The two most popular ones
proposed to date are probably dimensional regularization \cite{tHooft}
and the method of higher covariant derivatives \cite{HCD}. Dimensional
regularization works well for vector gauge theories, for which the
algebraic structure is not altered by a change in the number of
dimensions of spacetime. Unfortunately, when it comes to chiral gauge
theories, it is not clear \cite{CPM} \cite{GMR} whether it is possible
to consistently {\it define} dimensional regularization, the reason
being that the properties of chiral objects depend on the
dimensionality of spacetime and this conflicts somehow with the ideas
behind dimensional regularization.

The situation for the method of higher covariant derivatives is more
confusing. To describe it, we will restrict ourselves to QCD, the
theory we will be discussing here. As is well known, higher covariant
derivatives only provide a partial regularization, since they leave
one-loop divergences unregularized. To achieve full regularization, a
second regulator taking care of the unregularized one-loop divergences
must be introduced. Choosing such a regulator is not a simple issue,
since one would like to pick one that preserves gauge invariance and
that at the same time does not jeopardize what has been gained at two
and higher loops with the introduction of higher covariant
derivatives. Slavnov \cite{Slavnov} proposed in the seventies to use
as second regulator a certain combination of gauge invariant
Pauli-Villars determinants [see eq. (\ref{functional}) for their
expression]. Adopting his proposal, one ends up with a hybrid
regularization that combines higher covariant derivatives with
Pauli-Villars determinants of a certain form and that keeps the
dimension of spacetime at its physical value. We will call this
regularization prescription Slavnov's regularization and denote it by
SR. Keeping the dimensionality of spacetime unchanged and preserving
gauge invariance, SR looks like a good starting point to formulate a
suitable regularization method for chiral gauge theories.
Unfortunately, there is no agreement as for whether or not SR is a
consistent regularization method. On the one hand, there are claims
that (i) the Pauli-Villars determinants spoil regularization at two
and higher loops \cite{Warr} and (ii) that, even at one loop,
renormalization is inconsistent with gauge invariance \cite{Seneor}.
One the other, there are calls \cite{Faddeev} rebating these claims.

To settle down the dispute, and motivated by its potentiality for
chiral gauge theories, SR was used in ref. \cite{Nogo} to explicitly
regularize and renormalize Yang-Mills theory at one loop. There it was
proved that the Pauli-Villars determinants on which SR is based
generate unphysical logarithmic radiative corrections that modify the
beta function of the theory at one loop, giving for the latter an
unphysical value. The purpose of this paper is to show that these
unphysical corrections produce a violation of unitarity when the
Yang-Mills fields are coupled to fermionic matter, so that one has full
QCD.

\medskip

\noindent
{\it 2 Slavnov's regularization.} Let us start by briefly recalling
the basics of SR. We are interested in $N_\c\!$-coloured,
$N_\f\!$-flavoured QCD in four dimensions. Not to fall short of rigour
in the computation of Feynman integrals, we will work in Euclidean
space and recover Minkowski spacetime results by Wick rotating the
final results. In Euclidean space, QCD's classical action in a
covariant gauge $\partial A^a + \a\,b^a=0$ takes the form
\begin{equation}
  S = \idx \big( \caL_{\rm QCD} + \caL_{\rm GF} \big) \,,
\label{SQCD}
\end{equation}
where $\caL_{\rm QCD}$ and $\caL_{\rm GF}$ are given by
\begin{eqnarray}
  &{\ds \caL_{\rm QCD} =\, {1\over 4}\, F_{\m\n}^a \,F^{a\m\n}
      - \,\sum_{q=1}^{N_\f}\> \bar\psi_q\, \Big(\,i\parslash
      + ig \Aslash^a{}T^a\! - m_q\,\Big)\,\psi_q }&
\label{LQCD} \\[9pt]
  &{ \ds \caL_{\rm GF} =  -\,  {\a\over 2}~b^a b^a
      - b^a\,(\partial A^a) + \barc^a\, (\partial Dc)^a }&
\label{LGF}
\end{eqnarray}
and the notation is as follows. $\!A^a_\m$ denotes the gauge field,
$\bar\psi_q$ and $\psi_q$ the quark fields, $\barc^a$ and $c^a$
the Faddeev-Popov ghosts, $F^a_{\m\n} =
\partial_\m A^a_\n - \partial_\n A^a_\m + g f^{abc} A^b_\m A^c_\n\,$
the field strength and $D_\m^{ac} =\d^{ac} + g f^{abc} A^b_\m$ the
covariant derivative. $g$ is the coupling constant, $f^{abc}$ are the
structure constants of the gauge algebra, $T^a$ are the generators of
the gauge algebra in the fundamental representation, $m_q$ are the
fermion masses and $\a$ is the gauge-fixing parameter. The constants
$f^{abc}$ and the generators $T^a$ are normalized so that
$f^{acd}f^{bcd} \!=\!  N_\c\d^{ab}$ and ${\rm tr}(T^aT^b)={1\over
2}\d^{ab}.$ SR regularizes QCD in two steps. First, it introduces a
higher covariant derivative term and modifies the gauge-fixing part of
the action so that now one has
\begin{equation}
  S_\La = \idx \big( \caL_{\rm QCD} + \caL_{\rm HCD} + \caL'_{\rm GF}
        \big)\,,
\label{SLa}
\end{equation}
with
\begin{equation}
  \caL_{\rm HCD} = \, {1\over 4\La^4}\,
       (D^2 F_{\m\n})^a\, (D^2F^{\m\n})^a
\label{LHCD}
\end{equation}
and
\begin{equation}
\caL'_{\rm GF} = - \, {\a\over 2} \,b^a
               {1\over f^2(\partial^2\!/\La^2)}\,b^a
       - b^a \,(\partial A^a) + \barc^a\, (\partial D c)^a \>.
\label{LGFp}
\end{equation}
Here $\La$ is a mass and $f(\partial^2\!/\La^2)$ is a function that in
momentum space is given by
\begin{equation} f = 1 + {p^4\over \La^4} ~.
\label{choice}
\end{equation}
Some simple power counting shows that the only superficially divergent
1PI Feynman diagrams generated by $S_\La$ are the one-loop diagrams
contributing to the two, three and four-point 1PI Green functions of
the gauge field. Thus the modification of QCD's action along the lines
of eqs. (\ref{SLa})-(\ref{LGFp}) does not regularize the theory
completely, but leaves some one-loop divergences unregularized. Before
explaining Slavnov's idea to regularize the latter divergences, let us
mention that the choice of $f(\partial^2\!/\La^2)$ is somewhat
arbitrary. Here we have chosen it as in (\ref{choice}) so as to ensure
locality and make all $\a\!$-dependent contributions finite by power
counting. Taking \eg\ $f=1+(p^2/\La^2)$ also ensures locality but
leaves $\a\!$-dependent contributions unregularized at one loop (see
ref. \cite{Nogo} for a discussion of this point).

The second step in SR is to regularize the one-loop divergences
generated by $S_\La.$ Slavnov \cite{Slavnov} proposed to do this by
introducing Pauli-Villars determinants in the generating functional so
that it reads
\begin{equation} \begin{array}{l} {\ds
   Z\,[J,\chi,\zeta,\bar\zeta\,] \,= \int
     \,[\!dA]\, \,[\!db]\, \,[\!d\bar\psi]\, \,[\!d\psi]\,
     \,[\!d\barc]\, \,[\!dc]~
     e^{-\,\left( S_\La +\, S_{\rm source} \right)} }\\
\phantom{{\ds Z[\,J,\chi,\zeta,\bar\zeta\,] ~\, }} {\ds
     \,{\ss \times}\,\prod_{i=1}^I
             \left( \det {\BB A}_i \right)^{-\a_i/2}
             \left( \det {\BB C}_i \right)^{\a_i}
     \prod_{q=1}^{N_\f} \prod_{j=1}^{J_q}
             \left( \det {\BB F}_{qj} \right)^{\e_{qj}} \>.}
\end{array}
\label{functional}
\end{equation}
Here
\begin{displaymath}
   S_{\rm source} = \idx \bigg[ J^{a\m}A^a_\m + b^a \chi^a
     + \sum_{q=1}^{N_\f} \left(\, \bar\psi_q \zeta_q
     + \bar\zeta_q \psi_q\, \right) \bigg]
\end{displaymath}
is the source term coupling the fields $A^a_\m, ~b^a, ~\psi_q$ and
$\bar\psi_q$ to their external sources $J^{a\m}\!, ~\chi^a,
{}~\bar\zeta_q$ and $\zeta_q,$ and the determinants $\,\det{\BB A}_i,
{}~\det{\BB C}_i\,$ and $\det{\BB F}_{qj}$ are defined by
\begin{equation}
   \left( \det {\BB A}_i\,\right)^{-1/2} = \int\! \,[\!dA_i] ~ \d(DA_i)
        ~\exp \bigg\{ \!-{1\over2} \idx \!\idy  A^{a\m}_i(x)\,
        O^{~ab}_{i\,\m\n}(x-y)\, A^{b\m}_i(y) \bigg\}
\label{detA}
\end{equation}
\begin{equation}
   \det{\BB C}_i = \det \Big(\!-\!D^2\!+ M_i^2 \,\Big)
\label{detC}
\end{equation}
\begin{equation}
   \det{\BB F}_{qj} = \det \Big(\! i\parslash\,
          + g \Aslash^a{}T^a \!+ \m_{qj}\,\Big) \>.
\label{detF}
\end{equation}
The parameters $\a_i$ and $\e_{qj}$ are arbitrary real parameters
satisfying the conditions
\begin{eqnarray}
   & {\ds \sum_{i=1}^I \a_i + 1=0} & \label{PV1} \\
   & {\ds \sum_{j=1}^{J_q} \e_{qj} +1=0
     \qquad m_q^2 + \sum_{j=1}^{J_q} \,\e_{qj}\,\m_{qj}^2 = 0 \>,}
   & \label{PV2}
\end{eqnarray}
$M_i$ and $\m_{qj}$ are masses and the operator
$O^{~ab}_{i\,\m\n}(x-y)$ in eq. (\ref{detA}) is given by
\begin{displaymath}
   O^{~ab}_{i\,\m\n}(x-y) =
       {\d^2 S_\La\over\d A^a_\m(x)\,\d A^b_\n(y)}
        + M_i^2\,\d^{ab} g^{\m\n} \d(x-y) ~.
\end{displaymath}
Strictly speaking, Slavnov only considered pure Yang-Mills theory, so
he did not need to introduce determinants $\det{\BB F}_{qj}$ to
regularize the divergences generated by quarks running along internal
loops. We will see, anyway, that the determinants $\det{\BB F}_{qj}$
do not pose any problem and that the difficulties arise from the
determinants $\det{\BB A}_i.$ It is very easy to see \cite{Faddeev}
that $\det{\BB A}_i, ~\det{\BB C}_i$ and $\det{\BB F}_{qj}$ are gauge
invariant. This, together with the gauge invariance of $S_\La,$
implies that the functional $Z[J,\chi,\zeta,\bar\zeta]$ satisfies the
same BRS identities as the unregularized functional one would
construct starting from QCD's action $S$ in eq. (\ref{SQCD}).  It can
also be shown \cite{Slavnov} \cite{Nogo} that conditions (\ref{PV1})
and (\ref{PV2}) ensure that $Z[J,\chi,\zeta,\bar\zeta]$ generates
finite Green functions at one loop. So, all in all, one has a
generating functional which is manifestly gauge invariant and that
generates finite Green functions at one loop.

It is very important to understand the regularization mechanism of
one-loop divergences in $Z[J,\chi,\zeta,\bar\zeta].$ To this end, let
us consider the vacuum polarization tensor it generates. From the
action $S_\La,$ it receives the contributions of the diagrams in
Fig. 1. In addition there are the contributions from $\prod_i
(\det{\BB A}_i)^{-\a_i/2} (\det{\BB C}_i)^{\a_i} \prod_q \prod_j
\left(\det{\BB F}_{qj}\right)^{\e_{qj}}$ in the measure of the path
integral. Using the very same techniques as for ordinary Pauli-Villars
regularization of QED, it is very easy to see that conditions
(\ref{PV2}) imply that the product $\prod_q \prod_j \left(\det{\BB
F}_{qj}\right)^{\e_{qj}}$ regularizes the divergences in diagram
(1a). As for diagrams (1b)-(1d), Slavnov \cite{Slavnov} has argued
using formal path integral manipulations that violate locality that
the product $\prod_i (\det{\BB A}_i)^{-\a_i/2} (\det{\BB C}_i)^{\a_i}$
cancels the divergences in diagrams (1b)-(1d) provided eq. (\ref{PV1})
and the extra condition
\begin{displaymath}
  \sum_{i=1}^I \a_i M_i^2 =0
\end{displaymath}
are met. It has been shown \cite{Nogo} that, to check whether this is
actually the case within the framework of local regularization, an
extra regulator $\cR$ must be introduced. It turns out that after
introducing such a regulator and performing calculations at finite
$\cR,$ the 2-point divergences that arise in $\prod_i (\det{\BB
A}_i)^{-\a_i/2} (\det{\BB C}_i)^{\a_i}$ when $\cR\to 0$ cancel the
divergences that arise in diagrams (1b)-(1d) when $\cR\to 0,$ provided
only condition (\ref{PV1}) is satisfied. So indeed the sum of diagrams
(1a)-(1d) with the 2-point corrections from $\prod_i (\det{\BB
A}_i)^{-\a_i/2} (\det{\BB C}_i)^{\a_i} \prod_q \prod_j \left(\det{\BB
F}_{qj}\right)^{\e_{qj}}$ is finite if eqs. (\ref{PV1}) and
(\ref{PV2}) hold, but to see it without giving up locality,
an extra regulator $\cR$ is needed. The fact that one has to introduce
an extra regulator $\cR$ means strictly speaking that by itself SR
does not provide a local regularization of QCD.

Having a generating functional that generates finite Green functions
for finite values of the masses $\La,~M_i$ and $\m_{qj}$ is not
all. One has to devise a subtraction procedure that removes the
divergences associated to large values of the regulators $\La,~M_i$
and $\m_{qj},$ while preserving gauge invariance. There have been
claims in the past \cite{Seneor} that such a procedure does not exist
for pure Yang-Mills theory and that this is enough to kill SR. In
ref. \cite{Nogo}, however, it has been proved that this is not the
case and the most general subtraction procedure consistent with gauge
invariance for pure Yang-Mills theory has been given. Its
generalization to QCD being straightforward, we will not present
here. In what follows, we show that the functional
$Z[J,\chi,\zeta,\bar\zeta]$ generates unphysical contributions that,
after Wick rotation to Minkowski spacetime, spoil unitarity.

\medskip

\noindent {\it 3 Violation of unitarity.} Let us now come to Minkowski
spacetime, the correct framework to discuss unitarity. As is well
known, unitarity implies that the transition amplitude $T_{fi}$ for a
physical process $|i\rangle\to|f\rangle$ must satisfy the relation
\begin{equation}
  2 \>{\rm Im} \,T_{fi} = \sum_n\>
         (2\pi)^4\,\d^4(p_n-p_i)\> T^*_{nf}\, {T_{}}_{ni}\>,
\label{opth}
\end{equation}
where the sum is extended over all physical intermediate states
$|n\rangle$ connecting $|i\rangle$ with $|f\rangle$ and $p_n$ denotes
the momentum of the state $|n\rangle.$ Consider the process fermion,
antifermion going to fermion, antifermion. For this process, eq.
(\ref{opth}) takes at first order in perturbation theory the form
\begin{equation}
  2 \>{\rm Im} \,T_{1,\,\bar\f\f\to\bar\f\f} = \sum_n\>
     (2\pi)^4\,\d^4(p_n-p_i)\> {\vert T_{0,n\to\bar\f\f}\vert}^2\>.
\label{optical}
\end{equation}
The renormalized amplitude $T_1\equiv T_{1,\,\bar\f\f \to \bar\f\f}$
can be computed by first regularizing and then by subtracting the
divergences associated to the particular regulator one has used. This
way, the left-hand side becomes regularization and
subtraction-dependent. Actually, regularization-dependent only, since
different admissible subtraction schemes differ by finite local
renormalizations and these carry finite local radiative corrections
which do not reach the imaginary part of the transition amplitude. The
right-hand side, however, is regularization and
subtraction-independent, since it only involves the Feynman rules of
the unregularized theory. Hence eq. (\ref{optical}) can be viewed as a
necessary condition that the particular regularization and subtraction
prescriptions used to renormalize the theory must satisfy in order to
preserve unitarity. The idea of our proof of violation of unitarity by
SR is to compute ${\rm Im}\,T_1$ in any SR-based renormalization
scheme and see that it does not satisfy eq. (\ref{optical}). Now,
since dimensional regularization (DR) preserves unitarity, the
right-hand side in eq. (\ref{optical}) is equal to twice the imaginary
part of $T_1$ as computed in any DR-based renormalization scheme. Thus
eq. (\ref{optical}) can be replaced with
\begin{equation}
   2 \>{\rm Im} \,T_{1,\ss{\rm SR}} =
        2 \>{\rm Im} \,T_{1,\ss{\rm DR}} \>,
\label{SRDR}
\end{equation}
where the subscripts SR and DR refer to the regularization method used
to compute $T_1.$ In the sequel we show that eq. (\ref{SRDR}) does not
hold.

As is well known, the imaginary part of the amplitude $T_1$ receives
contributions from the diagrams depicted in Fig. 2, where all external
legs are on shell and the shadowed bubble stands for the vacuum
polarization tensor at one loop. To compute the contribution of these
diagrams to ${\rm Im}\,T_{1,\,\ss{\rm SR}},$ we proceed as
follows. We first calculate the renormalized contribution of every
diagram to the amplitude $T_{1,\ss\rm SR}$ in Euclidean space; once we
have done this, we Wick rotate to Minkowski spacetime; finally, we
take the imaginary part. As concerns the technical aspects of this
computation, we note that the calculation of the
renormalized contribution of any of the diagrams involved requires
computing its limit $\La,M_i,\m_{qj}\to\infty$ and subtracting
the divergences associated to this limit. The evaluation of such limit
is tedious but straightforward if one uses the techniques developed in
refs. \cite{GMR} and \cite{Nogo}. For simplicity, and since transition
amplitudes are independent of the gauge fixing parameter $\a,$ we will
work in the Feynman gauge $\a=1.$

We start by looking at diagrams (2a) and (2b). If we amputate the
external legs, we are left in both instances with a 1PI diagram whose
expression in Euclidean space in the Feynman gauge is
\begin{displaymath}
   G_4(p_1,p_2,p'_1;\La) = g^4\,T^aT^bT^bT^a \,
     \sum_{q=1}^{N_\f}\, I_q (\La) \>,
\end{displaymath}
where
\begin{displaymath}
  I_q(\La) = \idk
   {\ga_\m \> \big( \pslash_1 + \kslash - m_q \big) \,
    \ga_\n \, \ga_\r \> \big( \pslash'_1 + \kslash - m_q \big) \>
    \ga_\la \over [\, (p_1+k)^2+m_q^2\,] \>
                  [\, (p'_1+k)^2+m_q^2\,]} \>
    D^{\m\la}(p_1\!-\!p_2\!+\!k,\La) \> D^{\n\r}(k,\La)
\end{displaymath}
and
\begin{displaymath}
  D_{\m\n}(k,\La) = \,{\La^4\over k^4+\La^4} ~\bigg(\,
    {g_{\m\n}\over k^2} - {k_\m k_\n \over k^4 +\La^4} \bigg)
\end{displaymath}
is the gluon propagator for the action $S_\La$ in eq. (\ref{SLa}) with
$\a=1.$ Using the Lebesgue dominated convergence theorem, it is
straightforward to see that the $\La\to\infty$ limit of $I_q(\La)$ is
well defined and equal to
\begin{equation}
   \lim_{\La\to\infty} I_q(\La) = 4 \idk
   {\ga_\m \> \big( \pslash_1 + \kslash - m_q \big) \>
   \big( \pslash'_1 + \kslash - m_q \big) \> \ga^\m \over
   [\,(p_1+k)^2+m_q^2\,] \> [\,(p'_1+k)^2+m_q^2\,] \>
   (p_1\!-\!p_2\!+\!k)^2 \> k^2}~.
\label{G4ren}
\end{equation}
Hence no infinite renormalization is necessary. This is no surprise
since the unregularized 1PI four-fermion vertex $G_4$ at one loop is
finite by power counting and is given by the right hand-side in
eq. (\ref{G4ren}). This implies that the renormalized four-vertex
$G_4$ at one loop as computed with SR agrees with the renormalized
four-vertex $G_4$ as computed with any other acceptable regularization
method, and in particular with DR. It follows then, after Wick
rotating to Minkowski space and putting the external legs on shell,
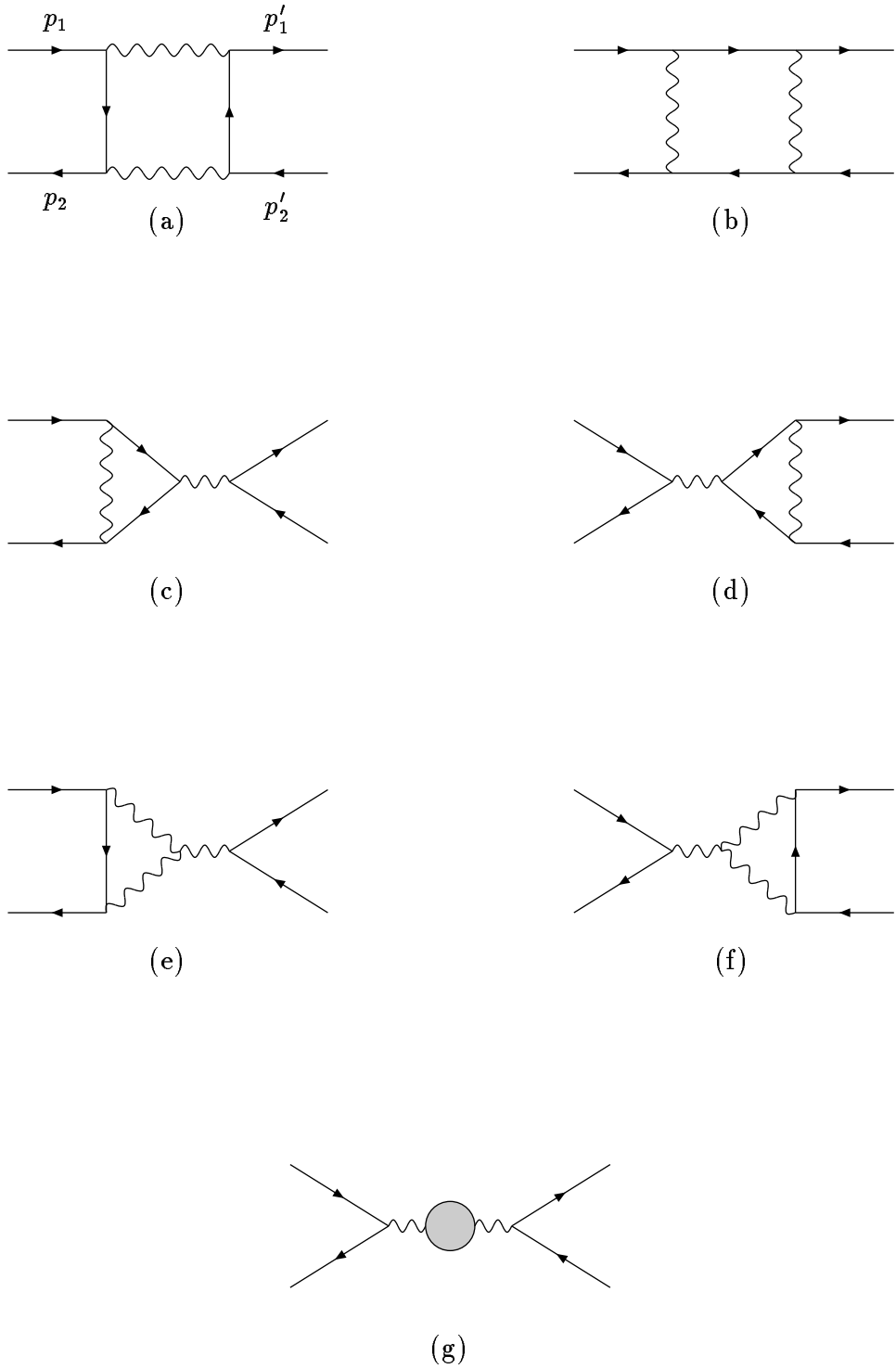
that the contribution of diagrams (2a) and (2b) to the imaginary part
of $T_1$ in any SR-based renormalization scheme is the same as in any
DR-based renormalization scheme:
\begin{equation}
  {\rm Im}\> T^{\,2a,2b}_{1,\,\ss{\rm SR}} =
      {\rm Im}\> T^{\,2a,2b}_{1,\,\ss{\rm DR}}\>.
\label{T2a}
\end{equation}

Next we move on to diagrams (2c) and (2d). They both have the 1PI
diagram in Fig. (3) as subdiagram. For the latter subdiagram,
SR gives in Euclidean space and in the Feynman gauge
\begin{displaymath}
  \gm^a_\m(p_1,p_2,\La) = ig^3\>{T^a\over 2N_c}\, \sum_{q=1}^{N_\f}
  \idk { \ga_\la \> \big(\pslash_2 + \kslash -m_q \big)\> \ga_\m \>
         \big( \pslash_1 + \kslash -m_q \big) \> \ga_\n \over
       [\,(p_2+k)^2 +m_q^2\,] \> [\,(p_1+k)^2 +m_q^2\,]} \>
       D^{\la\n}(k,\La) ~.
\end{displaymath}
Using the techniques in ref. \cite{GMR} to compute the large-$\!\La$
limit, and the results in ref. \cite{Aoki} to rewrite the
contributions that do not vanish in this limit, we obtain
\begin{equation}
  \lim_{\La\to\infty} \>
  \gm^a_\m(p_1,p_2,\La) = \,{ig^3\over 8\pi^2} \>{T^a\over 2N_c}~
      \sum_{q=1}^{N_\f}\, \bigg\{ \ga_\m \,\bigg[ \,
      {1\over 2}\>\ln\!\bigg({\La^2\over m_q^2}\bigg) + v_0\,\bigg]
      + F^{\,\rm fin}_\m\, (m_q,p_1,p_2)\, \bigg\} \>,
\label{G3reg}
\end{equation}
where $v_0$ is a numerical constant and the finite part is given by
\begin{equation}
\begin{array}{l} {\ds
  F^{\,\rm fin}_\m\,(m_q,p_1,p_2) = \int_0^1 \!dx \int_0^1
    \! dy \, y\, \ln\bigg[{m_q^2 \over D(x,y)}\bigg] } \\[10pt]
\phantom{ {\ds F^{\,\rm fin}_\m\,(m_q,_1,p_2)~}} {\ds
  + \int_0^1\!dx \int_0^1\!dy ~
    {m_q^2 \ga_\m - 2m_q \,\big(p_1+p_2-2y\,\bar p\big)_\m
        + \big(\pslash_1-y\bar\pslash\big)\,\ga_\m\,
          \big(\pslash_2-y\bar\pslash\big)  \over D(x,y) }} ~.
\end{array}
\label{finite}
\end{equation}
Here $D(x,y)$ and $\bar p^\m$ stand for
\begin{displaymath}
  D(x,y) =
       m_q^2 + (p_2 -p_1)^2\,x\,(1-x) + y\,(1-y)\,\bar p^2
\end{displaymath}
and
\begin{displaymath}
  \bar p^\m = (1-x)\,p^\m_1 + x\,p_2^\m \>.
\end{displaymath}
We see that $\gm^a_\m(p_1,p_2,\La)$ diverges as $\La\to\infty.$ To
remove the divergence, we perform the most general subtraction
compatible with gauge invariance and obtain
\begin{equation}
  \gm^a_{\m,\,{\rm ren}}(p_1,p_2,\s) = \,{ig^3\over 8\pi^2} \>
    {T^a\over 2N_c}~ \sum_{q=1}^{N_\f}\, \bigg\{ {1\over 2}\>
    \ga_\m \,\bigg[ - \ln\!\bigg({m_q^2\over\s^2}\bigg) +
     \,v\bigg({m_q^2\over \s^2}\bigg) \bigg] +
    F^{\,\rm fin}_{\m}\,(m_q,p_1,p_2) \bigg\} \>,
\label{G3ren}
\end{equation}
$\s$ being the renormalization mass scale and $v\,(m_q^2/\s^2)$
a finite function that does not depend on the momenta and that is
only restricted by BRS invariance (in a minimal scheme, it would be
zero). Let us now recall what DR gives. In DR, instead of
eq. (\ref{G3reg}) one has
\begin{displaymath}
  \lim_{\ee\to 0} \>
  \gm^a_\m(p_1,p_2;\ee,\n) = \,{ig^3\over 8\pi^2} \>{T^a\over 2N_c}~
      \sum_{q=1}^{N_\f}\, \bigg\{ {1\over 2}\,\ga_\m \,
      \bigg[ \,{1\over \ee} +
      \,\ln\!\bigg({\n^2\over m_q^2}\bigg) + v_0\,\bigg]
      + F^{\,\rm fin}_\m\, (m_q,p_1,p_2)\, \bigg\} \>,
\end{displaymath}
where $\n$ is the dimensional regularization mass scale and $v_0$ is a
constant different from that in eq. (\ref{G3reg}). After
renormalization, one obtains the same expression as in
eq. (\ref{G3ren}), modulo finite local radiative corrections. This
implies, after Wick rotating to Minkowski space and replacing the
subdiagram in Fig. (3) with its renormalized expression, that the
imaginary part of diagrams (2c) and (2d) is the same for SR-based
renormalization schemes as for DR-based renormalization schemes:
\begin{equation}
  {\rm Im}\> T^{\,2c,2d}_{1,\,\ss{\rm SR}} =
      {\rm Im}\> T^{\,2c,2d}_{1,\,\ss{\rm DR}}\>.
\label{T2c}
\end{equation}
Proceeding analogously, it is easy to see that the same holds true
for diagrams (2e) and (2f):
\begin{equation}
  {\rm Im}\> T^{\,2e,2f}_{1,\,\ss{\rm SR}} =
      {\rm Im}\> T^{\,2e,2f}_{1,\,\ss{\rm DR}}\>.
\label{T2e}
\end{equation}

Let us finally look at diagram (2g), and let us concentrate on the
one-loop vacuum polarization tensor $\Pi^{ab}_{\m\n}(k)$ hidden in it.
The latter is made of two contributions. First there is the
contribution (we will denote it with a prime) from diagram (1a) and
the determinants $\det{\BB F}_{qj}$ that regularize it. Its expression
can be computed following the very same steps as for the vacuum
polarization tensor in Pauli-Villars regularized QED
\cite{Itzykson}. After some calculations, we get
\begin{equation}
  \Pi^{\prime}{}^{\,ab}_{\m\n}\,(k,\m_{qj}) = -\,
    {g^2 \over 16\pi^2} \,\d^{ab}\,{2\over 3}\,
    \sum_{q=1}^{N_\f}\, \bigg[ \sum_{j=1}^{J_q}\,\e_{qj}\,
           \ln\!\bigg({m_q^2\over\m_{qj}^2}\bigg) +\,
           h \bigg({m_q^2\over k^2}\bigg) +\,\pi'_0 \,
           \bigg]\, \big( k_\m k_\n - k^2 g_{\m\n} \big) \>,
\label{Piferm}
\end{equation}
where $h(x)$ is the function
\begin{displaymath}
   h(x) = 4x +\, (1-2x)\;\sqrt{1+4x}~
        \ln\!\bigg({ {\sqrt{1+4x}+1}\over{\sqrt{1+4x}-1}}\bigg)
\end{displaymath}
and $\pi'_0$ is a numerical constant. Then there is the contribution
from diagrams (1b)-(1d) and from the Pauli-Villars determinants
$\det{\BB A}_i$ and $\det{\BB C}_i.$ This contribution (we will denote
it with a double prime) is the same as for pure Yang-Mills theory, and
its $\La,M_i\to\infty$ limit has been computed for arbitrary $\a$ in
ref. \cite{Nogo}. Borrowing the results from there, we have that in
the Feynman gauge
\begin{equation}
  \Pi^{\prime\prime}{}^{\,ab}_{\m\n}\,(k,\La,M_i) = \,
  {g^2N_\c\over 16\pi^2}\,
  \d^{ab}\bigg[ A_2\,\ln\!\bigg({k^2\over\La^2}\bigg) +\, B_2
  \,\sum_{i=1}^I \,\a_i \ln\!\bigg({k^2\over M_i^2}\bigg) +\,\pi''_0
  \,\bigg] \,\big(k_\m k_\n -k^2g_{\m\n}\big) \>,
\label{PiYM}
\end{equation}
where
\begin{equation}
  A_2-B_2={11\over 6}
\label{A2B2}
\end{equation}
and $\pi''_0$ is another numerical constant. The actual values of
$A_2$ and $B_2$ depend on the way the masses $\La$ and $M_i$ are sent
to infinity. For example, sending $\La$ to infinity while keeping
$M_i$ finite, and taking in the result $M_i\to\infty$ gives different
$A_2$ and $B_2$ as proceeding the other way around. The difference
$A_2-B_2$ is, however, independent of the path followed to approach
$\La,M_i\to\infty$ and is always given by eq. (\ref{A2B2}). Summing
the contributions (\ref{Piferm}) and (\ref{PiYM}), performing the most
general subtraction compatible with gauge invariance, and Wick
rotating to Minkowski space, we obtain for the renormalized vacuum
polarization tensor
\begin{equation}
  \Pi^{ab}_{\m\n,\,{\rm ren}}\,(k,\s) =\,{g^2\over 16\pi^2}~
     \Pi_{\ss \rm SR} (k^2\!,\s) ~ \d^{ab} \,
     ( k_\m k_\n - k^2g_{\m\n}) \>,
\label{Piren}
\end{equation}
where
\begin{equation}
  \Pi_{\ss \rm SR} (k^2\!,\s) =\,{11N_\c\over 6}\;
     \ln\!\bigg(\!\!-\!{k^2\over \s^2}\bigg) - {2\over 3}\,
     \sum_{q=1}^{N_\f} \,\bigg[\, h\bigg(\!\!-\!{m_q^2\over k^2}\bigg)
     +\, \ln\!\bigg({m_q^2\over\s^2}\bigg)
     + \,\pi\bigg({m_q^2\over\s^2}\bigg) \bigg] \>,
\label{PiSR}
\end{equation}
$\s$ is the subtraction point and $\pi(m_q^2/\s^2)$ is an
arbitrary real function carrying local finite radiative corrections
constrained only by BRS invariance. For momenta $k^\m$ such that
$k^2\!>\!0,$ the vacuum polarization tensor picks an imaginary part
since
\begin{equation}
  {\rm Im}\> \Pi_{\ss \rm SR} (k^2\!,\s) = \,
     {11N_\c \over 6}\> \theta(k^2)
     + {2\over 3}\> \sum_{q=1}^{N_\f} \, \theta(k^2\!-4m_q^2) \,
       \bigg(1+{2m_q^2\over k^2}\bigg)\, \sqrt{1 -{4m_q^2\over k^2}} ~.
\label{ImSR}
\end{equation}
Let us compare this with the DR result. We recall that in any DR-based
subtraction scheme, one has in the Feynman gauge that
\begin{equation}
  {\rm Im}\> \Pi_{\ss {\rm DR}}(k^2\!,\s) = \, {5 N_\c\over 3}
                                                          \>\theta(k^2)
     + {2\over 3}\> \sum_{q=1}^{N_\f} \, \theta(k^2\!-4m_q^2) \,
       \bigg(1+{2m_q^2\over k^2}\bigg)\, \sqrt{1 - {4m_q^2\over k^2}} ~ .
\label{ImDR}
\end{equation}
We see that the coefficient in front of $\theta(k^2)$ is different
from that in eq. (\ref{ImSR}). As explained in ref. \cite{Nogo}, the
difference ${11\over 6}-{5\over 3}= {1\over 6}$ is originated by the
Pauli-Villars determinants $\det{\BB A}_i.$ It is plain now that the
imaginary part of the renormalized contribution
\begin{displaymath}
  T^{\,2g}_1 = g^2\, \Big[\bar v(p_2)\ga^\m T^a u(p_1) \Big] \,
   \Pi(k^2\!,\s)\> {1\over k^4} \;
   \big( k_\m k_\r - k^2 g_{\m\r} \big) \,
   \Big[ \bar u(p'_1) \ga^\r T^a v(p'_1) \Big]
\end{displaymath}
of diagram (2g) to the amplitude $T_1$ is not the same for SR-based
renormalization schemes as for DR-based schemes. In other words,
\begin{equation}
  {\rm Im}\> T^{\,2g}_{1,\,\ss{\rm SR}} \neq
      {\rm Im}\> T^{\,2g}_{1,\,\ss{\rm DR}}\>.
\label{T2g}
\end{equation}
Putting together eqs. (\ref{T2a}), (\ref{T2c}), (\ref{T2e}) and
(\ref{T2g}), we have that
\begin{displaymath}
{\rm Im}\> T_{1,\,\ss{\rm SR}} \neq
      {\rm Im}\> T_{1,\,\ss{\rm DR}}\>,
\end{displaymath}
as we wanted to prove.

\medskip

\noindent{\it 4. Conclusion.}  At this point we can draw the following
conclusions:

\noindent
(i) The regularization method proposed by Slavnov violates unitarity,
the violation being produced by the Pauli-Villars determinants
$\det{\BB A}_i$ that the method chooses. Let us try to gain some
intuition of why this is so. Assume that we naively switch off the
regulators in the regularized path integral in eq. (\ref{functional}),
that is to say, that we send the masses $\La,~M_i$ and $\m_{qj}$ to
infinity. Then we should recover the unregularized QCD path
integral. However, this is not the case \cite{Nogo} \cite{Asorey}. To
see the latter, we rescale \cite{Asorey} the Pauli-Villars field
$A^a_{i\,\m} \to M^{-1}\,A^a_{i\,\m}$ in $(\det{\BB A}_i)^{-1/2},$
take the limit $\La,M_i\to\infty,$ exponentiate $\d(DA_i)$ and
integrate over $d^4\!x$ once OBby parts. This leaves us with
\begin{displaymath}
   (\det{\BB A}_i)^{-1/2} \sim \int \,[\!dA_i]\, \,[\!db_i]\,
   \exp\bigg\{ - {1\over 2} \idx \Big( A_i^2 + 2\,A_i D b_i
   \Big) \bigg\} \quad {\rm as~} \quad\La,M_i\to\infty \>.
\end{displaymath}
Completing the square in the exponent and performing the integral
yields $(\det D^2)^{-1/2}.$ Since each $(\det{\BB A}_i)^{-1/2}$ is
exponentiated to the power $\a_i$ and the $\a_i\!$'s satisfy
eq. (\ref{PV1}), we obtain a factor $(\det D^2)^{1/2}.$ As for the
determinants $\det{\BB C}_i$ and $\det{\BB F}_{qj},$ it is
straightforward to see that their naive limits $M_i\to\infty$ and
$\m_{qj}\to\infty$ give unity. Thus taking the naive
$\La,M_i,\m_{qj}\to\infty$ limit in $Z[J,\chi,\zeta,\bar\zeta]$ yields
the unregularized QCD path integral plus an extra $(\det D^2)^{1/2}.$
This extra determinant introduces propagating degrees of freedom that
couple to the gluon field through the covariant derivative and which
are not present in QCD's action. In other words, SR modifies QCD even
at the tree level. Obviously the properties of the modified QCD are
not the same as those of the true QCD. In the light of this, it is
very easy to understand SR's violation of unitarity. What
$T_{1,\,\ss{\rm SR}}$ is really standing for is the transition
amplitude $T_1$ for the modified theory. By the cutting rules of 't
Hooft and Veltman \cite{Diagrammar}, ${\rm Im}\,T_{1,\,\ss{\rm SR}}$
will receive contributions from the new propagating degrees of
freedom. Hence there is no way ${\rm Im}\,T_{1,\,\ss{\rm SR}}$ will
agree with the imaginary part of $T_{1,\,\ss \rm QCD}\equiv
T_{1,\,\ss\rm DR}.$ All this discards SR as an acceptable
regularization method.

\noindent
(ii) Note that the diagrams whose regularization only involves $\La,$
namely diagrams (2a) to (2f), give the correct contribution to
$T_{1,\,\ss\rm QCD}$ in the $\La\to\infty$ limit. This indicates that
the higher covariant derivatives terms in eq. (\ref{LHCD}) by
themselves do not cause problems, in agreement with \cite{nonmult}.
The question that remains open is to supplement higher covariant
derivatives with a suitable local regularization that preserves gauge
invariance manifestly. Let us recall in this regard that for a local
regularization method to be such, it must provide integrals over loop
momenta which are finite by power counting (this is what local
regularization is about). If a prescription does not provide this
finiteness by power counting, it should not be called a local
regularization; not even in the event that the various divergent
contributions from different divergent Feynman integrals cancel among
themselves when the latter are properly defined through yet another
regularization.

\bigskip\bigskip

%
\def\section{\subsection}

\end{document}